# Hybrid Surface Plasmon Polaritons (HSPPs) in Plasma-based Elliptical Waveguides with Graphene Layers

**Mohammad Bagher Heydari [1,*], Morteza Mohammadi Shirkolaei [2], Majid Karimipour [3]**

[1,*] School of Electrical Engineering, Iran University of Science and Technology (IUST), Tehran, Iran
[2] Department of Electrical Engineering, Shahid Sattari Aeronautical University of Science and Technology, Tehran, Iran
[3] Department of Electrical Engineering, Arak University of Technology, Arak, Iran

[*]Corresponding author: mo_heydari@alumni.iust.ac.ir

**Abstract:** In this article, tunable surface plasmon polaritons (SPPs) in graphene-based elliptical waveguides containing gyro-electric layers are investigated. The general structure has an elliptical cross-section, where each gyro-electric layer is surrounded by two graphene layers. The DC magnetic bias is applied on the z-axis. As a special case, a new plasma-based elliptical structure with double-layer graphene is studied in this paper. The figure of merit (FOM) for this waveguide can be varied by changing the magnetostatic bias and the chemical doping. At the frequency of 40 THz, the FOM of 139 for this waveguide is reported for the B= 1 T and $\mu_c$=0.9 eV. Ability to adjust and tune the propagating properties of SPPs in hybrid graphene-plasma elliptical structures can be exploited for the design of new plasmonic components in the THz spectral region.

## 1. Introduction

The optical conductivity of graphene, a tunable parameter that can be adjusted by electrostatic bias (or chemical doping) and magnetostatic bias, plays the main role in the design and fabrication of graphene-based plasmonic devices in the THz frequencies [1-3]. This key feature makes graphene a fascinating material in the THz region in which many various devices have been presented and reported in the literature such as couplers [4-6], filters [7-9], resonators [10-12], circulators [13-16], waveguides [17-26], sensing [27-32], imaging [33, 34], hyperbolic structures [35-43], and polarization converters [44]. Graphene-based waveguides have various structures such as planar [18, 26, 45-75], cylindrical [56, 76-81], and elliptical structures [25, 82-84].

To enhance the performance of graphene-based nanostructures, one familiar way is by integrating graphene with smart materials, such as chiral materials, nonlinear magnetic materials, and Kerr-type materials, which will help the designer to easily change and control the propagating features of SPPs. For instance, a new promising platform has been proposed for sensing applications by the hybridization of graphene with chiral materials in [26]. Another important point is the usage of this integration on a suitable structural platform to increase the degrees of freedom. For instance, two geometric factors in elliptical structures, i.e. short and long axis, can assist the designer in adjusting the modal properties of nanostructures, as reported in [82, 83, 85-87]. Therefore, the hybridization of graphene with a gyroelectric medium in an elliptical structure can be interesting due to various tunability parameters. Gyroelectric materials have many fascinating applications in the literature such as isolators [88]. The gyroelectric material used in this paper is InSb. InSb is one of the fascinating semiconductors in magneto-plasmonics since it provides some remarkable properties such as low effective mass [89], strong cyclotron frequencies at small magnetic fields [90], and the Faraday effect [91]. This paper suggests new mathematical expressions for graphene-plasma waveguides. The presented model is fast compared to computational methods because they discretize the waveguide to give a solution while the proposed model does not require any discretization. In addition, the model gives closed-form relations for all propagating SPPs.

Our multi-layer structure containing graphene layers can be useful for the design and implementation of novel THz components. The propagating properties of our waveguide can be modified via external magnetic bias and the chemical potential of graphene layers, which can allow one to obtain high values of the effective index and FOM. Moreover, more degrees of freedom due to having two geometric factors in elliptical structures, i.e. short and long axis, can help the designer in adjusting the modal properties of these structures. The remainder of the article is organized as follows. In section 2, the general waveguide will be introduced and then an analytical model will be obtained by embarking on Maxwell's equations. In section 3, after validation of the model by comparing the analytical results with simulation ones for an elliptical nano-wire, the propagation features of a new elliptical gyro-electric waveguide with double-layer graphene will be studied. Finally, the article is concluded in section 4.

## 2. The General Structure and its Analytical Model

In Fig. 1, the cross-section of the general waveguide has been depicted. In this structure, each graphene layer is placed among two different plasma layers. A DC magnetic bias is applied on the z-axis. To show the generality of the proposed waveguide defined in fig. 1, we have utilized various blue colors for different gyro-electric materials. In other words, the intensities of blue colors in this figure are only used to represent the variety of gyro-electric layers.

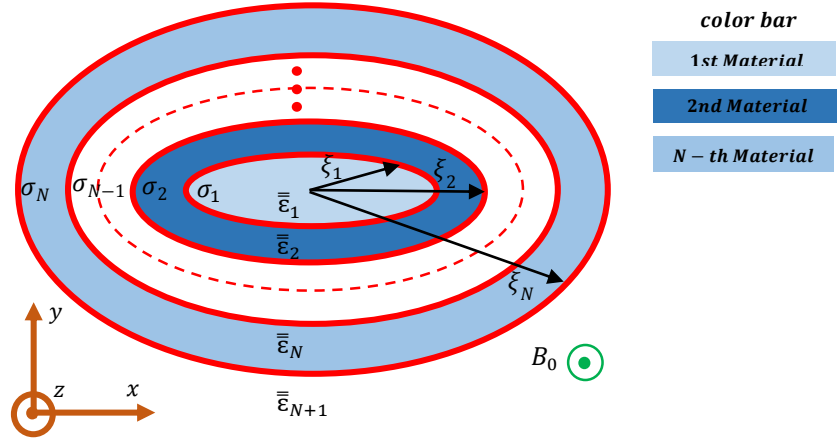

**Fig. 1.** The cross-section schematic of the general structure. The structure is infinite in the z-direction.

In the *N*-th layer, the permittivity tensor of the gyro-electric medium is expressed as [92]:

$$\bar{\bar{\varepsilon}}_N = \varepsilon_0 \begin{pmatrix} \varepsilon_N & j\varepsilon_{a,N} & 0 \\ -j\varepsilon_{a,N} & \varepsilon_N & 0 \\ 0 & 0 & \varepsilon_{\|,N} \end{pmatrix} \qquad (1)$$

With the following elements [92]:

$$\varepsilon_N = \varepsilon_{\infty,N}\left(1 - \frac{\omega_{p,N}^2(\omega + j\upsilon_N)}{\omega\left[(\omega + j\upsilon_N)^2 - \omega_{c,N}^2\right]}\right) \qquad (2)$$

$$\varepsilon_{a,N} = \varepsilon_{\infty,N}\left(\frac{\omega_{p,N}^2 \omega_c}{\omega\left[(\omega + j\upsilon_N)^2 - \omega_c^2\right]}\right) \qquad (3)$$

$$\varepsilon_{\parallel,N} = \varepsilon_{\infty,N}\left(1 - \frac{\omega_{p,N}^2}{\omega(\omega+j\upsilon_N)}\right) \tag{4}$$

In (2)-(4), $\upsilon_N$ and $\varepsilon_{\infty,N}$ are the collision rate and the background permittivity, respectively. Furthermore,

$$\omega_{p,N} = \sqrt{\frac{n_s e^2}{\varepsilon_0 \varepsilon_{\infty,N} m^*}} \tag{5}$$

$$\omega_c = \frac{eB_0}{m^*} \tag{6}$$

are known as the plasma and cyclotron frequencies of the plasma medium. In (5)-(6), $e, m^*$ and $n_s$ are the charge, effective mass, and density of the carriers, respectively. It should be mentioned that the conductivity of the $N$-th layer can be modeled by the following relation [1]:

$$\sigma_N(\omega,\mu_{c,N},\Gamma_N,T) = \frac{-je^2}{4\pi\hbar}Ln\left[\frac{2|\mu_{c,N}| - (\omega - j2\Gamma_N)\hbar}{2|\mu_{c,N}| + (\omega - j2\Gamma_N)\hbar}\right] + \frac{-je^2 K_B T}{\pi\hbar^2(\omega - j2\Gamma_N)}\left[\frac{\mu_{c,N}}{K_B T} + 2Ln\left(1 + e^{-\mu_{c,N}/K_B T}\right)\right] \tag{7}$$

In relation (7), $\Gamma_N$ is the scattering rate, $T$ is the temperature, and $\mu_{c,N}$ is the chemical potential for the $N$-th layer [1].

The proposed structure has an elliptical cross-section and thus it should be analyzed in elliptical coordinates $(\xi,\eta)$ [93]. In the frequency domain, Maxwell's equations inside the $N$-th can be written as follows [94]:

$$\nabla \times \mathbf{E} = -j\omega\mu_N \mathbf{H} \tag{8}$$

$$\nabla \times \mathbf{H} = j\omega\bar{\bar{\varepsilon}}_N \cdot \mathbf{E} \tag{9}$$

By defining the following operator in the elliptical coordinates ($f$ is the semi-focal length of the ellipse):

$$\nabla_\perp^2 = \frac{1}{f\sqrt{\cosh^2\xi - \cos^2\eta}}\left(\frac{\partial^2}{\partial^2\xi} + \frac{\partial^2}{\partial^2\eta}\right) \tag{10}$$

The z-components of the electromagnetic fields inside the gyro-electric layer are obtained:

$$\left(\nabla_\perp^2 + \frac{\varepsilon_{\parallel,N}}{\varepsilon_N}\frac{\partial^2}{\partial z^2} + (k_0^2 \varepsilon_{\parallel,N}\mu_N)\right)E_{z,N} + k_0\mu_N\left(\frac{\varepsilon_{a,N}}{\varepsilon_N}\right)\frac{\partial H_{z,N}}{\partial z} = 0 \tag{11}$$

$$\left(\nabla_\perp^2 + \frac{\partial^2}{\partial z^2} + (k_0^2 \varepsilon_{\perp,N}\mu_N)\right)H_{z,N} - k_0\varepsilon_{\parallel,N}\left(\frac{\varepsilon_{a,N}}{\varepsilon_N}\right)\frac{\partial E_{z,N}}{\partial z} = 0 \tag{12}$$

and,

$$\varepsilon_{\perp,N} = \varepsilon_N - \frac{\varepsilon_{a,N}^2}{\varepsilon_N} \tag{13}$$

Let us assume that SPPs propagate in the z-axis and hence the electromagnetic fields are written as follows ($m$ is the order of Mathieu functions):

$$H_{z,N}(\xi,\eta,z) = \int_{-\infty}^{+\infty}\sum_{m=0}^{\infty}e^{jk_z z}H_{z,N}^m(\xi,\eta;q)\,dk_z \tag{14}$$

$$E_{z,N}(\xi,\eta,z) = \int_{-\infty}^{+\infty}\sum_{m=0}^{\infty}e^{jk_z z}E_{z,N}^m(\xi,\eta;q)\,dk_z \tag{15}$$

Substitute (14)-(15) into (11)-(12) to achieve the following equation:

$$s^4 + G_{1,N}s^2 + G_{2,N} = 0 \tag{16}$$

where

$$G_{1,N} = -k_0^2 \mu_N \left( \varepsilon_{\parallel,N} + \varepsilon_{\perp,N} \right) + \left( \frac{\varepsilon_{\parallel,N}}{\varepsilon_N} + 1 \right) k_z^2 \tag{17}$$

$$G_{2,N} = \left( k_0^2 \varepsilon_{\parallel,N} \mu_N - \frac{\varepsilon_{\parallel,N}}{\varepsilon_N} k_z^2 \right) \left( k_0^2 \varepsilon_{\perp,N} \mu_N - k_z^2 \right) - k_0^2 k_z^2 \mu_N \varepsilon_{\parallel,N} \left( \frac{\varepsilon_{a,N}}{\varepsilon_N} \right)^2 \tag{18}$$

are coefficients of equation (16). Let us define the following parameters ($S_{2N-1}, S_{2N}$ are the roots of (16) for the N-th layer):

$$q_{2N-1} = \frac{f^2}{4} S_{2N-1} \tag{19}$$

$$q_{2N} = \frac{f^2}{4} S_{2N} \tag{20}$$

Now, for different regions, the roots of characteristics equations can be considered as follows (*N* shows the number of the layer and *i* expresses the index of the roots for that layer):

$$q = \begin{cases} q_1, q_2 & i = 1, 2 \; ; \; \textit{for first layer} \\ \ldots & \ldots \\ q_{2N-1}, q_{2N} & i = 2N-1, 2N \; ; \; \textit{for N-th layer} \\ q_{2N+1}, q_{2N+2} & i = 2N+1, 2N+2 \; ; \; \textit{for N+1-th layer} \end{cases} \tag{21}$$

The z-components of fields for various regions can be written as:

$$H_z^m(\xi,\eta;q) = \begin{cases} A_{m,1,1} Je_m(\xi,q_1) ce_m(\eta,q_1) + A'_{m,1,1} Jo_m(\xi,q_1) se_m(\eta,q_1) + \\ A_{m,2,1} Je_m(\xi,q_2) ce_m(\eta,q_2) + A'_{m,2,1} Jo_m(\xi,q_2) se_m(\eta,q_2) & \xi < \xi_1 \\ \ldots \\ ce_m(\eta,q_{2N-1}) \left[ A_{m,2N-1,N} Je_m(\xi,q_{2N-1}) + B_{m,2N-1,N} Ne_m(\xi,q_{2N-1}) \right] + \\ se_m(\eta,q_{2N-1}) \left[ A'_{m,2N-1,N} Jo_m(\xi,q_{2N-1}) + B'_{m,2N-1,N} No_m(\xi,q_{2N-1}) \right] + \\ ce_m(\eta,q_{2N}) \left[ A_{m,2N,N} Je_m(\xi,q_{2N}) + B_{m,2N,N} Ne_m(\xi,q_{2N}) \right] + \\ se_m(\eta,q_{2N}) \left[ A'_{m,2N,N} Jo_m(\xi,q_{2N}) + B'_{m,2N,N} No_m(\xi,q_{2N}) \right] & \xi_{N-1} < \xi < \xi_N \\ B_{m,2N+1,N+1} Ne_m(\xi,q_{2N+1}) ce_m(\eta,q_{2N+1}) + \\ B'_{m,2N+1,N+1} No_m(\xi,q_{2N+1}) se_m(\eta,q_{2N+1}) + \\ B_{m,2N+2,N+1} Ne_m(\xi,q_{2N+2}) ce_m(\eta,q_{2N+2}) + \\ B'_{m,2N+2,N+1} No_m(\xi,q_{2N+2}) se_m(\eta,q_{2N+2}) & \xi > \xi_N \end{cases} \tag{22}$$

$$E_z^m(\xi,\eta;q) = \begin{cases} T_{1,1}\begin{pmatrix} A'_{m,1,1}Je_m(\xi,q_1)ce_m(\eta,q_1)+ \\ A_{m,1,1}Jo_m(\xi,q_1)se_m(\eta,q_1) \end{pmatrix}+ \\ T_{2,1}\begin{pmatrix} A'_{m,2,1}Je_m(\xi,q_2)ce_m(\eta,q_2)+ \\ A_{m,2,1}Jo_m(\xi,q_2)se_m(\eta,q_2) \end{pmatrix} & \xi < \xi_1 \\ \ldots \\ T_{2N-1,N}\,ce_m(\eta,q_{2N-1})\begin{bmatrix} A'_{m,2N-1,N}Je_m(\xi,q_{2N-1})+ \\ B'_{m,2N-1,N}Ne_m(\xi,q_{2N-1}) \end{bmatrix}+ \\ T_{2N-1,N}\,se_m(\eta,q_{2N-1})\begin{bmatrix} A_{m,2N-1,N}Jo_m(\xi,q_{2N-1})+ \\ B_{m,2N-1,N}No_m(\xi,q_{2N-1}) \end{bmatrix}+ \\ T_{2N,N}\,ce_m(\eta,q_{2N})\begin{bmatrix} A'_{m,2N,N}Je_m(\xi,q_{2N})+ \\ B'_{m,2N,N}Ne_m(\xi,q_{2N}) \end{bmatrix}+ \\ T_{2N,N}\,se_m(\eta,q_{2N})\begin{bmatrix} A_{m,2N,N}Jo_m(\xi,q_{2N})+ \\ B_{m,2N,N}No_m(\xi,q_{2N}) \end{bmatrix} & \xi_{N-1}<\xi<\xi_N \\ T_{2N+1,N+1}\begin{bmatrix} B'_{m,2N+1,N+1}Ne_m(\xi,q_{2N+1})ce_m(\eta,q_{2N+1})+ \\ B_{m,2N+1,N+1}No_m(\xi,q_{2N+1})se_m(\eta,q_{2N+1}) \end{bmatrix}+ \\ T_{2N+2,N+1}\begin{bmatrix} B'_{m,2N+2,N+1}Ne_m(\xi,q_{2N+2})ce_m(\eta,q_{2N+2})+ \\ B_{m,2N+2,N+1}No_m(\xi,q_{2N+2})se_m(\eta,q_{2N+2}) \end{bmatrix} & \xi > \xi_N \end{cases}$$

(23)

In (23), $T_{i,N}$ are:

$$T_{i,N} = \frac{q_i^2 - k_0^2 \varepsilon_{\perp,N}\mu_N + k_z^2}{-jk_0 k_z \varepsilon_{\|,N}\left(\dfrac{\varepsilon_{a,N}}{\varepsilon_N}\right)} \qquad i = 2N-1, 2N \quad ; \quad N = 1,2,\ldots$$

(24)

It should be noted that $Je_m(\xi,q), Jo_m(\xi,q), Ne_m(\xi,q), No_m(\xi,q)$ are radial Mathieu functions of the first and second kind, respectively and $ce_m(\xi,q), se_m(\xi,q)$ are angular Mathieu functions, respectively [93]. Now, the transverse components of electromagnetic fields are obtained:

$$\begin{pmatrix} E_{\xi,N} \\ H_{\xi,N} \end{pmatrix} = \frac{1}{f\sqrt{\cosh^2\xi - \cos^2\eta}}\left[\bar{\bar{Q}}_N^{Pos}\frac{\partial}{\partial\xi}\begin{pmatrix} E_{z,N} \\ H_{z,N} \end{pmatrix} + j\bar{\bar{Q}}_N^{Neg}\frac{\partial}{\partial\eta}\begin{pmatrix} E_{z,N} \\ H_{z,N} \end{pmatrix}\right] \qquad (25)$$

$$\begin{pmatrix} E_{\varphi,N} \\ H_{\varphi,N} \end{pmatrix} = \frac{-j}{f\sqrt{\cosh^2\xi - \cos^2\eta}}\left[\bar{\bar{Q}}_N^{Neg}\frac{\partial}{\partial\xi}\begin{pmatrix} E_{z,N} \\ H_{z,N} \end{pmatrix} + j\bar{\bar{Q}}_N^{Pos}\frac{\partial}{\partial\eta}\begin{pmatrix} E_{z,N} \\ H_{z,N} \end{pmatrix}\right] \qquad (26)$$

Where

$$\bar{\bar{Q}}_N^{Pos} = \frac{1}{2}\left[\frac{1}{-k_z^2 + k_0^2 \varepsilon_{+,N}\mu_{+,N}}\begin{pmatrix} jk_z & -\omega\mu_0\mu_{+,N} \\ \omega\varepsilon_0\varepsilon_{+,N} & jk_z \end{pmatrix} + \frac{1}{-k_z^2 + k_0^2 \varepsilon_{-,N}\mu_{-,N}}\begin{pmatrix} jk_z & \omega\mu_0\mu_{-,N} \\ -\omega\varepsilon_0\varepsilon_{-,N} & jk_z \end{pmatrix}\right]$$

(27)

$$\bar{\bar{Q}}_N^{Neg} = \frac{1}{2}\left[\frac{1}{-k_z^2 + k_0^2 \varepsilon_{+,N}\mu_{+,N}}\begin{pmatrix} jk_z & -\omega\mu_0\mu_{+,N} \\ \omega\varepsilon_0\varepsilon_{+,N} & jk_z \end{pmatrix} - \frac{1}{-k_z^2 + k_0^2 \varepsilon_{-,N}\mu_{-,N}}\begin{pmatrix} jk_z & \omega\mu_0\mu_{-,N} \\ -\omega\varepsilon_0\varepsilon_{-,N} & jk_z \end{pmatrix}\right]$$

(28)

are the Q-matrices used in (25) and (26). In addition,

$$\varepsilon_{\pm,N} = \varepsilon_N \pm \varepsilon_{a,N} \qquad (29)$$

is used in (27)-(28). By utilizing the boundary conditions,

$$E_{z,N} = E_{z,N+1}, E_{\eta,N} = E_{\eta,N+1}$$
$$H_{z,N+1} - H_{z,N} = -\sigma E_{\eta,N}, H_{\eta,N+1} - H_{\eta,N} = \sigma E_{z,N} \qquad N = 1,2,3,... \qquad (30)$$

Thus, the final matrix for the structure of Fig. 1 is achieved:

$$\bar{\bar{S}}_{8N,8N} \cdot \begin{pmatrix} A_{m,1,1} \\ A'_{m,1,1} \\ A_{m,2,1} \\ A'_{m,2,1} \\ ... \\ B_{m,2N+1,N+1} \\ B'_{m,2N+1,N+1} \\ B_{m,2N+2,N+1} \\ B'_{m,2N+2,N+1} \end{pmatrix}_{8N,1} = \begin{pmatrix} 0 \\ 0 \\ 0 \\ 0 \\ ... \\ 0 \\ 0 \\ 0 \\ 0 \end{pmatrix}_{8N,1} \qquad (31)$$

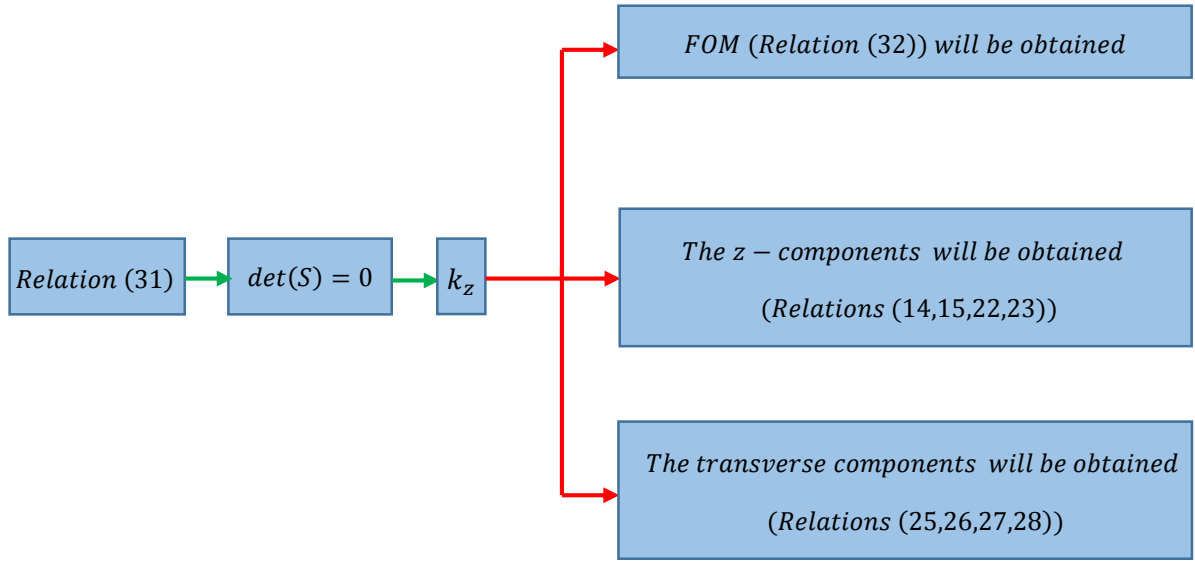

Fig. 2. The block diagram of the proposed mathematical model.

Now, setting $det(\bar{\bar{S}}) = 0$ will obtain the propagation constant and other propagating features such as effective index and figure of merit (FOM). The FOM can be defined by the following relation:

$$FOM = \frac{1}{\lambda_g \cdot \text{Im}[n_{eff}]} \xrightarrow{\lambda_g = \frac{2\pi}{\text{Re}[n_{eff}]}} FOM = \frac{\text{Re}[n_{eff}]}{2\pi \cdot \text{Im}[n_{eff}]} \xrightarrow{n_{eff} = \frac{k_z}{k_0}} FOM = \frac{\text{Re}[k_z]}{2\pi \cdot \text{Im}[k_z]} \qquad (32)$$

In (32), we have defined a "benefit-to-cost" ratio utilizing the guided wavelength ($\lambda_g$), and the imaginary part of the effective index (Im[$n_{eff}$]). The inverse wavelength ($\frac{1}{\lambda_g}$) shows the confinement of propagating SPPs and the imaginary part of the effective index (Im[$n_{eff}$]) indicates the losses. This defined FOM can be utilized for practical applications where a small wavelength ($\lambda_g$) is important, such as nano-waveguides. To better clarify the relationship between FOM and the analytical expressions obtained in this section, Fig. 2 is depicted. This figure represents the mathematical procedure of the proposed model in which the FOM and the electromagnetic field distributions are achieved by starting from relation (31).

## 3. Results and Discussions

In this section, a new graphene-based gyro-electric waveguide will be studied, as a special case of the general structure. In all results, the temperature is $T = 300\ K$, the graphene thickness is supposed to be $\Delta = 0.33\ nm$, and the scattering rate is $\Gamma = 2 \times 10^{12}\ rad/s$. To be brief, only the first two plasmonic modes are investigated here.

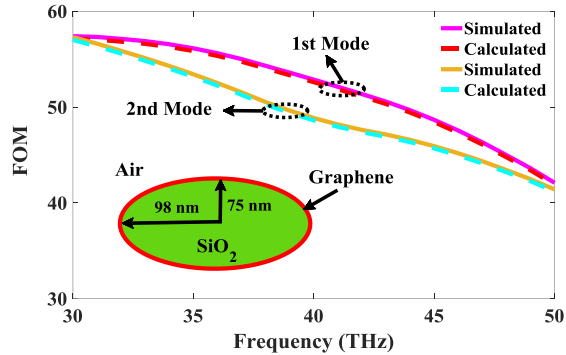

**Fig. 3.** Dependence of FOM on frequency for a graphene nano-wire. The dashed lines are analytical results prepared by the proposed model and the solid lines are simulation results prepared by COMSOL software.

To show the accuracy of the analytical relations outlined in the previous section, the simulation results of FOM with analytical ones for a graphene nano-wire are compared. As seen in Fig. 3, a graphene layer is located on an elliptical dielectric, with a permittivity of 2.09 ($\varepsilon_{SiO_2} = 2.09$) and the geometrical parameters are $a = 98\ nm, b = 75\ nm$. In this figure, the chemical potential is supposed to be 0.45 eV ($\mu_c = 0.45\ eV$). One of the most important parameters for the investigation of nano-waveguides is FOM [95]. In Fig. 3, the analytical and simulation results of FOM for the first two propagating modes in a graphene-based nano-wire are reported. In the simulation, the analysis mode of COMSOL software is utilized to obtain the propagation features. The mode analysis is a powerful method to compute the propagation constants and investigate the mode characteristics. The numerical method is the finite element method (FEM), a popular numerical method that discretizes the space into small points, and the perfectly matched layers (PML) have been used as the boundary conditions. The mesh sizes of the structure are $\Delta x = \Delta y = \Delta z = 0.2\ nm$ in all directions. It should be noted that the analytical model proposed in this article is more accurate and faster than any computational method (such as FEM) because it gives closed-form relations for electromagnetic distributions and propagation features, without discretization of the structure. It can be observed from this figure that FOM for the 1st mode is higher than for the 2nd mode. This happens because the effective index for the 1st mode is higher than the 2nd mode for a certain propagation loss. Moreover, a good agreement between simulation and analytical results is seen, which confirms the proposed analytical model. In what follows, the analytical results of FOM (prepared by the analytical model) are reported.

Consider a new elliptical gyro-electric waveguide with double-layer graphene, as illustrated in Fig. 4. In this figure, its cross-section is only shown. Here, n-type InSb is chosen as gyroelectric layer due to its large magneto-optical response at mid-infrared region. Moreover, it is common in graphene plasmonics that graphene layers should be deposited on a familiar substrate. In our calculations, to present a practical design, $SiO_2$-Si layers are chosen as a support substrate. The permittivity of Si and $SiO_2$ layers are assumed to be 11.9 and 2.09 ( $\varepsilon_{Si} = 11.9$, $\varepsilon_{SiO_2} = 2.09$), respectively. The gyro-electric layer is the n-type InSb with these parameters [96]: $\varepsilon_\infty = 15.68$, $m^* = 0.022 m_e$, $n_s = 1.07 \times 10^{17}/cm^3$, $\nu = 0.314 \times 10^{13} s^{-1}$ and $m_e$ is the electron's mass. The chemical potential of the graphene is 0.45 eV ($\mu_c = 0.45\ eV$). Furthermore, the geometrical parameters are $a_1 = 75\ nm, b_1 = 55\ nm, a_2 = 95\ nm, b_2 = 80.15\ nm, a_3 = 105\ nm, b_3 = 91.78\ nm$.

In Fig. 5, we have depicted the plasmonic properties of the two first propagating modes, i.e. the first and second modes, as a function of frequency for various values of magnetic bias ($B_0 = 0, 1, 2$ T). The propagation length has been defined as: $L_{Prop} = \lambda_g/4\pi Im[n_{eff}]$. One can see from this figure that the real part of effective index increases with the frequency increment while the propagation length decreases. Altering magnetic bias can drastically increase the real part of effective index of the first mode (as seen in Fig. 5 (a) for $B_0 = 2$ T) while it has a negligible effect on the propagation features of the second mode.

Fig. 6 illustrates the FOM of the elliptical gyro-electric waveguide as a function of frequency for the first two modes for various magnetic fields. As seen in this figure, the FOM for both propagating modes decreases with the frequency increment. It occurs because the propagation loss increases with the frequency increment and thus FOM decreases. It is evident from this figure that FOM can be enhanced by increasing the applied bias, especially for the 1st mode. Indeed, the magnetic field increment has an insignificant influence on the FOM of the 2nd mode. It happens because the most energy is concentrated on the short axes of the waveguide which means that the electromagnetic distribution is almost uniform for the second mode and thus altering the magnetic field will not change the FOM of this mode. Fig. 7 illustrates the electric field distributions of the first and second modes. The frequency is 40 THz. It can be seen from this figure that the energy distribution is non-uniform for the first mode and has been concentrated in a small area, i.e. graphene-InSb-graphene area.

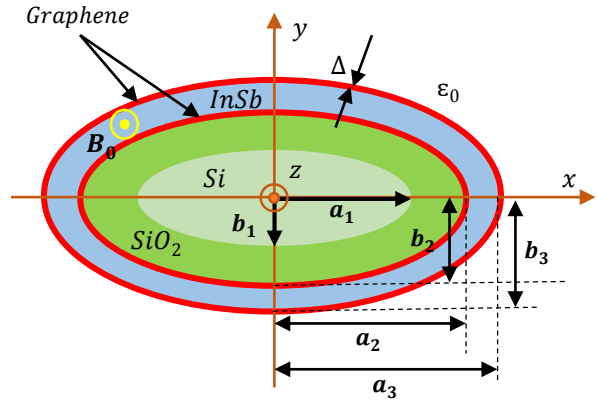

**Fig. 4.** The cross-section of the elliptical gyro-electric waveguide with double-layer graphene.

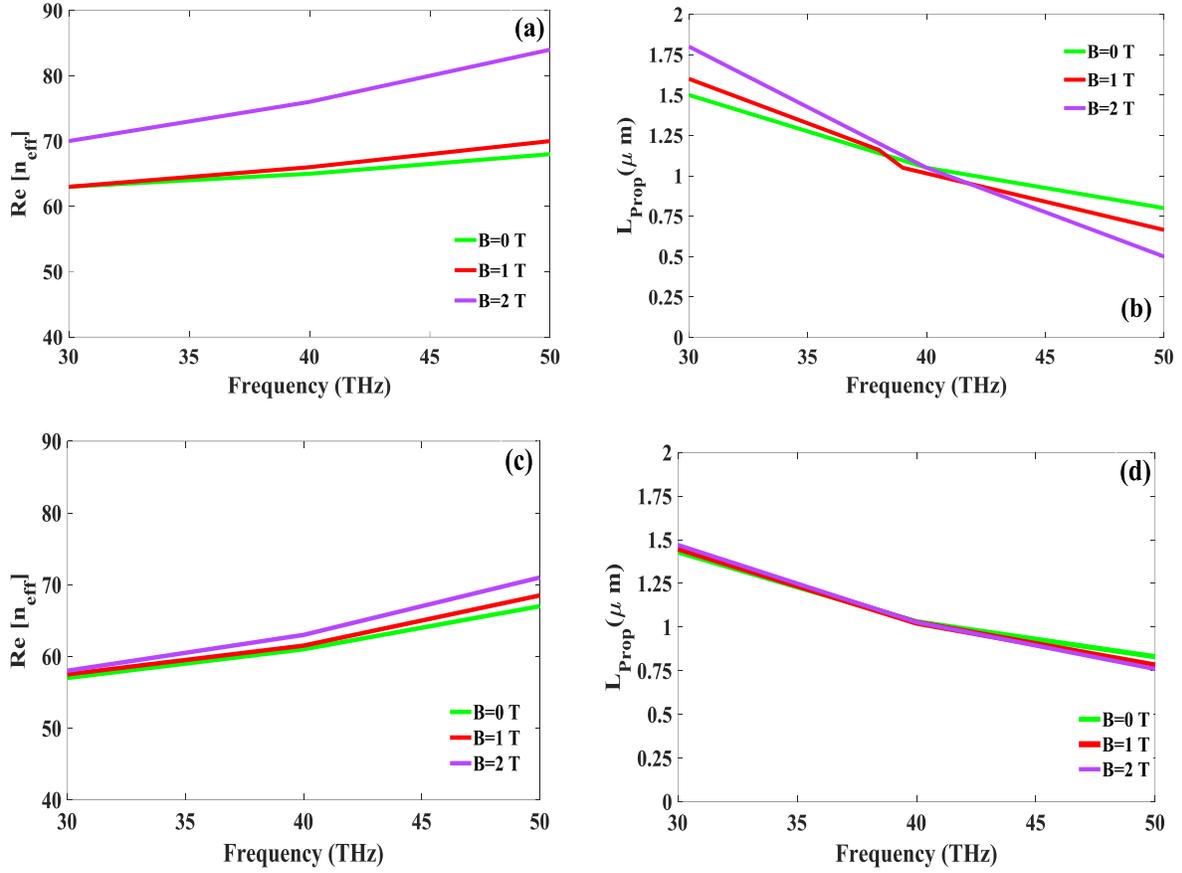

**Fig. 5.** The real part of effective index and propagation length for various magnetic biases ($B_0 = 0, 1, 2$ T) for: (a) and (b) First mode. (c) and (d) Second mode.

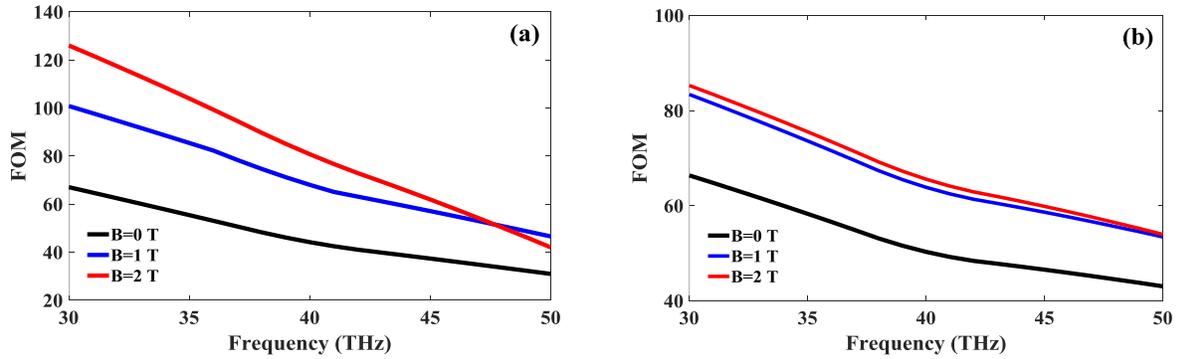

**Fig. 6.** FOM versus frequency for several magnetic biases ($B_0 = 0, 1, 2$ T) for: (a) 1st mode, (b) 2nd mode.

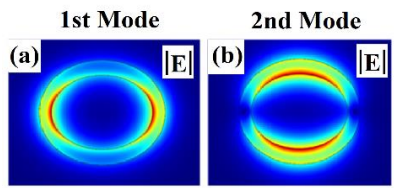

**Fig. 7.** The electric field distribution for: (a) The first mode, (b) The second mode. The frequency is supposed to be 40 THz.

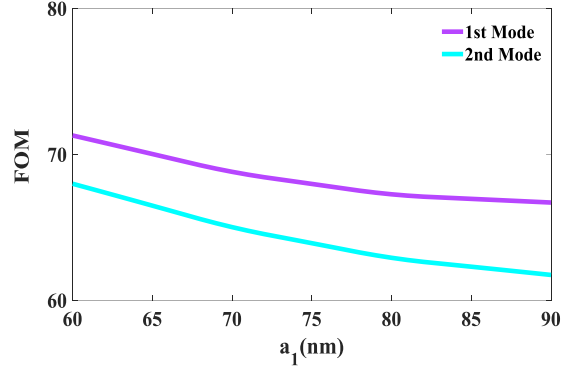

**Fig. 8.** Dependence of FOM on the long axis ($a_1$) of Si for the 1st and 2nd modes.

Now, the effect of the variations of geometrical factors on FOM is considered. Fig. 8 shows FOM as a function of the long-axis ($a_1$) of the Si layer. It is supposed that the frequency is 40 THz. The DC magnetic bias is 1 T and the chemical potential is 0.45 eV. It can be observed that FOM decreases as the long-axis increases. In addition, the 1st mode has a better FOM than the 2nd mode, as seen in fig. 8. The slope of FOM variations for the 2nd mode is greater than the 1st mode, as the long-axis increases. In fig. 9, FOM as a function of the short-axis ($b_1$) of the Si layer has been depicted for the first two modes. FOM of the 1st mode is higher than 2nd mode for the range of $b_1 < 68\ nm$. As the short-axis increases ($b_1 > 68\ nm$), the FOM for the two modes becomes equal.

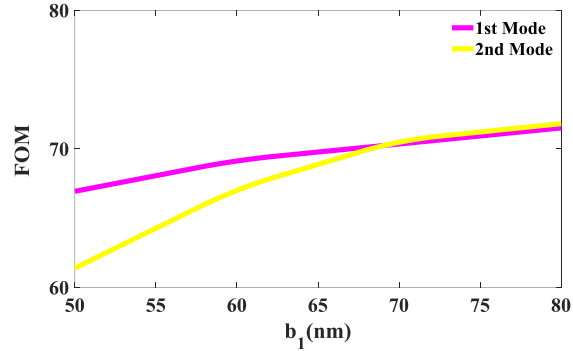

**Fig. 9.** Dependence of FOM on the short axis ($b_1$) of Si for the 1st and 2nd modes.

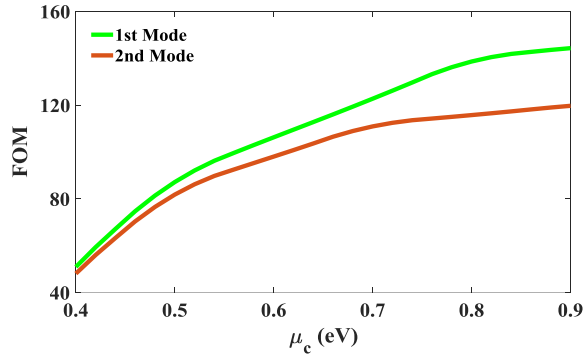

**Fig. 10.** FOM versus the chemical potential at the frequency of 40 THz.

As a final point, it is worthwhile to be noted that the elliptical gyro-electric waveguide with double-layer graphene is a tunable structure in which its propagation properties can be changed via chemical potential. Fig. 10 demonstrates the dependence of FOM on chemical potential. As mentioned before, the graphene layers in the structure have similar parameters. By increasing the chemical potential of graphene layers, the FOM increases. It happens because the change of chemical potential alters the surface conductivity of graphene layers. However, the slope of FOM variations is greater for the 1st mode, especially for the range of $\mu_c > 0.7\ eV$. One can observe that a high value of FOM is obtainable for this waveguide. For instance, FOM reaches 139 for $\mu_c = 0.9\ eV$ for the 1st mode. It should be noted that the fabrication procedures of the proposed waveguide are similar to graphene nano-wires, which are discussed in [97-102]. To explain briefly, an InSb nanowire is placed on a graphene layer by micromanipulation technique [103] and then a graphene layer is placed on graphene-InSb layers. After that, Silica and Silicon layers are deposited on the obtained structure. Finally, to eliminate the waveguide from the substrate, a tapered fiber is applied [97-99]. Some experiments have been reported for graphene coating on $SiO_2$ nanowires which the reader can refer to them to better understand the fabrication process of graphene-based nanowires [104].

## 4. Conclusion

In this article, a new theoretical model is reported for general elliptical structures with graphene layers. FOM of a graphene-based elliptical waveguide, forming graphene-InSb-graphene-$SiO_2$-Si layers, was calculated and studied at the frequency range of 30-50 THz. Only the first two SPP modes were studied in this paper. The proposed waveguide could support tunable, non-reciprocal SPPs in which their modal properties are varied by chemical potential and the external magnetic field. The author believes that the integration of graphene layers and gyro-electric materials can open new research areas in tunable non-reciprocal devices in the mid-infrared region.

**Declarations**

**Ethics Approval:** Not Applicable.

**Consent to Participate:** Not Applicable.

**Consent for Publication:** Not Applicable.

**Funding:** The authors received no specific funding for this work.

**Conflicts of Interest/ Competing Interests:** The authors declare no competing interests.

**Availability of Data and Materials:** Not Applicable.

**Code availability:** Not Applicable.

**Authors' Contributions:** M. B. Heydari proposed the main idea of this work and performed the analytical modeling. M. Mohammadi Shirkolaei conducted the numerical simulations and wrote the manuscript. M. Karimipour analyzed the results and reviewed the paper.